\documentclass[reprint, superscriptaddress, amsmath,amssymb, aps,prl,floatfix,nobalancelastpage]{revtex4-1}
\usepackage{graphicx}
\usepackage{dcolumn}
\usepackage{bm}
\usepackage{hyperref}
\usepackage[mathlines]{lineno}
\usepackage{textcomp}
\usepackage{gensymb}
\usepackage[super]{nth}
\begin{document}


\title{Coherent Control of the Non-instantaneous Nonlinear Power-law Response in Resonant Nanostructures}

\author{Eyal Bahar}
\email{Eyalb@mail.tau.ac.il}
\author{Uri Arieli}%
\author{Michael Mrejen}
\author{Haim Suchowski}
\affiliation{School of Physics and Astronomy, Tel-Aviv University, Tel-Aviv 6779801, Israel}
\affiliation{Center for Light-Matter Interaction, Tel-Aviv University, Tel-Aviv 6779801, Israel}%
\date{\today}
\begin{abstract}
We experimentally demonstrate coherent control of the nonlinear response of optical second harmonic generation in resonant nanostructures beyond the weak-field regime. Contrary to common perception, we show that maximizing the intensity of the pulse does not yield the strongest nonlinear power-law response. We show this effect emerges from the temporally asymmetric photo-induced response in a resonant mediated non-instantaneous interaction. We develop a novel theoretical approach which captures the photoinduced nonlinearities in resonant nanostructures beyond the two photon description and give an intuitive picture to the observed non-instantaneous phenomena. 
\end{abstract}
\setlength{\parskip}{0pt}
\maketitle
Nanostructures (NS) have revolutionized light matter interaction allowing for on demand control of unique optical \cite{avayu2017composite,yang2007enhanced}, electrical \cite{rybka2016sub,newson2008coherently} and mechanical properties \cite{liu2012micromachined}, both in linear and nonlinear regimes \cite{camacho2016nonlinear,li2017nonlinear,hasan2018nanoparticles,catherine2017gold}. In the past decade, much research has been performed on the optical nonlinearity of NSs emerging from their energy confinement \cite{krasnok2018nonlinear} and geometrical architecture \cite{o2015predicting,husu2012metamaterials} contributing in both their single and collective responses \cite{fernandez2017unrelenting,michaeli2017nonlinear,valev2009plasmonic}. Commonly enhanced by resonant NSs, the photoinduced nonlinear interaction in NSs has been mostly studied within the framework of the instantaneous response of these materials \cite{liu2016resonantly,kauranen2012nonlinear,amendola2017surface}, meaning that the nonlinear medium interacts simultaneously with all interacting waves. While this instantaneous picture has provided a model describing the observations of rich nonlinear phenomena, it does not capture the full nonlinear dynamical response, which is fundamentally non-instantaneous. 
This is apparent in current research frontiers, where the study of the ultrafast, out-of-equilibrium, electronic dynamics in NSs has gained much attention \cite{lang2018ultrafast,roller2017hotspot,putnam2017optical,lazzarini2017linear,roloff2017light,born2016ultrafast,huang2016special,shcherbakov2015ultrafast,harutyunyan2015anomalous,brongersma2015plasmon}. However, the non-instantaneous contribution inherent to  \textit{resonant} interaction in these systems has been so far mostly overlooked.
\par The non-instantaneous contribution inherent to nonlinear resonant dynamics is well known in atomic and molecular systems, which is of particular importance in multiphoton processes \cite{shapiro2003principles}. Enhancement by orders of magnitude of electronic transitions in atomic systems \cite{dudovich2001transform,levin2015coherent} as well as for large organic molecules \cite{lozovoy2003multiphoton,oron2002quantum}, has been enabled by spectrally shaping the pulse to be compatible with the non-instantaneous response in resonant mediated interactions, via coherent control schemes. However, for resonant NSs, applying such pulse shaping methods to enhance nonlinear processes have been so far limited, since these require the interacting pulse spectrum to be much broader than the resonant linewidth, which is typically not fulfilled for NSs. Therefore, pulse shaping has been mostly shown for controlling the \textit{linear} response in plasmonic systems \cite{stockman2002coherent,brinks2013plasmonic,onishi2013spatiotemporal,volpe2010deterministic,kojima2016control} or for multicolor second harmonic generation (SHG) imaging \cite{accanto2016resonant}. As the resonant mediated non-instantaneous process also contributes to the power-law response, it is expected that this fundamental characteristic of the nonlinear dynamical response will play a major role in nonlinear interaction. Yet, to date, no experimental work has been performed demonstrating control over the non-instantaneous nonlinear power-law response in resonant NSs.
\par In this letter we experimentally demonstrate coherent control of the non-instantaneous nonlinear power-law response in resonant NSs. We show that, counter-intuitively, the highest peak intensity does not yield the strongest optical nonlinearity, where the nonlinear response decreases significantly below the standard quadratic response. Furthermore, when approaching the strong field regime, we reveal an asymmetric temporal evolution accompanied with an unconventional decrease in the nonlinear power-law response, deviating from 2 to 1.6. Finally, we devise a theoretical framework and provide an intuitive picture for the non-instantaneous effects. Our proposed model, based on a resonant three-level system \cite{yariv1991optical}, solved to the fourth order in a perturbative expansion, captures non-instantaneous resonant phenomena beyond the weak-field two-photon description. In this framework, phase differences in the excitations of the localized surface plasmonic resonance (LSPR), captured as phase differences between second and fourth order contributions, lead to a dynamically rich nonlinear response which we observe experimentally. These effects are not captured in traditional nonlinear models including the well-known Miller rule \cite{boyd2003nonlinear}, which has shown great accuracy in describing non-resonant nonlinearities in bulk media. Though our experimental observations were focused on plasmonic NSs, we believe the resonant three-level model is critical in analyzing strong light-matter interaction with any resonant process in NSs, including excitonic, electronic or phononic excitations in semiconductor NSs and 2D materials.
   \begin{figure}[ht]
\centering\includegraphics[width=\linewidth]{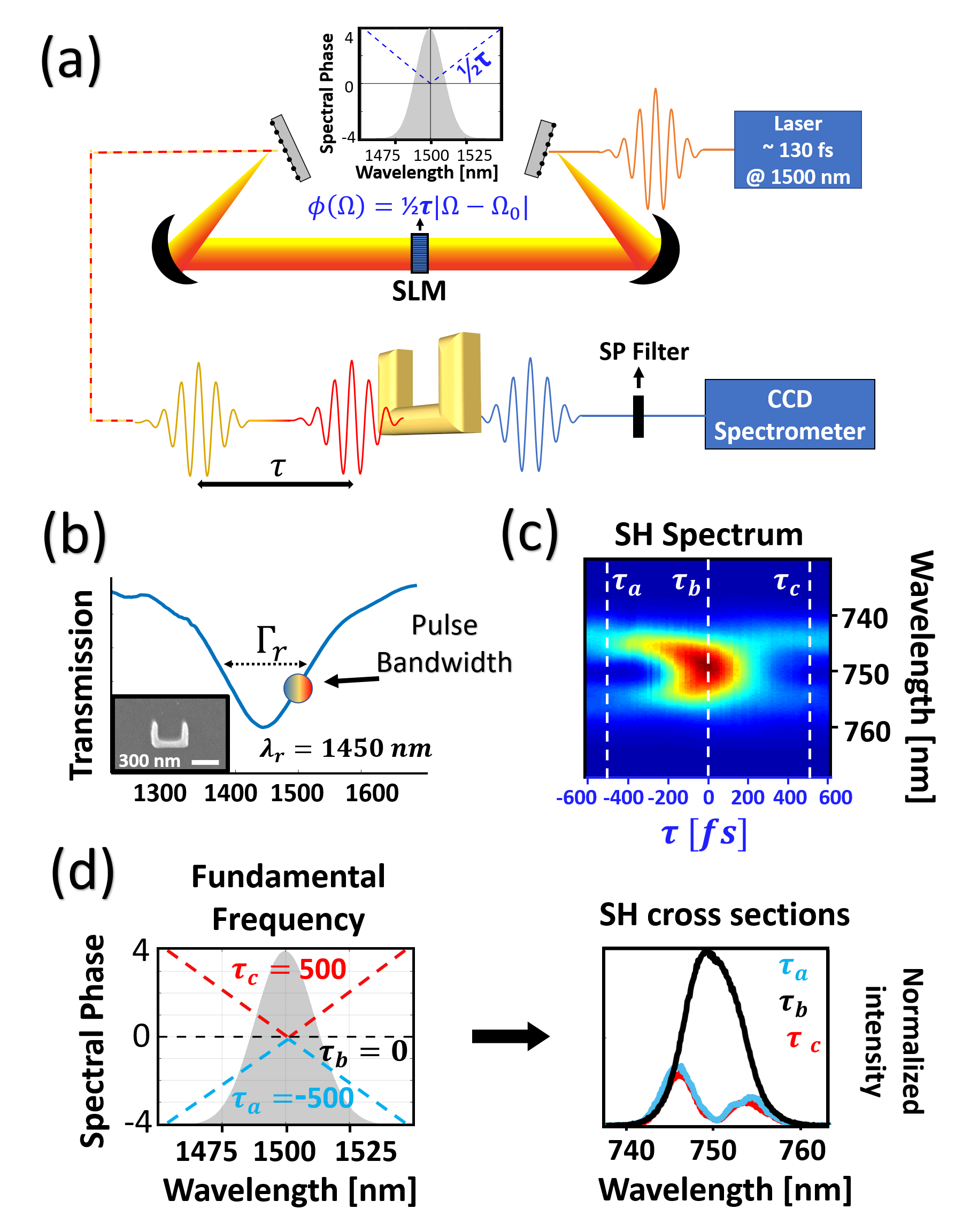}
\caption{Coherent control of SHG in plasmonic NSs. (a) The experimental apparatus is composed of a spatial light modulator (SLM) used as a spectral phase mask. Light passes through a 4f pulse shaper SLM in a double-pass configuration. The coherently altered pulse interacts with a gold NS, SHG is collected. (b) NS transmission measurement showing the LSPR wavelength and linewidth. Inset: The measured image of a SRR by a scanning electron microscope. (c) A normalized map of the measured SHG spectrum as a function of the spectral phase is placed as example of a weak-field reference when observing the power-law response map. The horizontal axis of the SH map is the delay between the spectrally split pulses, corresponding to the slope of an absolute value phase function centered about the central frequency of the pulse. (d) Cross sections of the SH map showing the SHG for large delays and with no delay (TL pulse).}
\label{fig:atmosphere}
	\end{figure}
\par In our experimental demonstration of the non-instantaneous response, we use gold NSs in a split ring resonator (SRR) configuration, which is the most common geometrical structure with inherent SHG. In all experiments we have used a 130-fs laser fixed at 1500 nm focused on a SRR array with LSPR in the vicinity of the lasers central wavelength [see Figure \ref{fig:atmosphere}(a),(b)]. For the coherent control experiment, we have assembled a pulse shaper consisting of a spatial light modulator (SLM) based in a double-pass 4f setup. When performing SLM-based experiments in NSs by a \textit{single-pass} set-up, one should be aware of the potential spatial shifts and distortions of the beam due to spatiotemporal couplings in the 4f-system. The \textit{double-pass} configuration, was found to be indeed crucial to obtain accurate SLM-based measurements \cite{brinks2011beating}. In our experiments, we apply a variable absolute valued spectral phase mask spectrally splitting the incoming pulse. By Fourier transform, the magnitude of the slope of the absolute valued phase function, $\tau/2 $, is directly related to the temporal interval between the spectrally separated pulses. The sign of the slope, positive or negative, corresponds to the order of arrival of the spectrally separated pulses where positive dictates longer wavelengths arrive first. This enables to perform an SLM based, semi-degenerate coherent pump-probe SHG process and create a map emphasizing the non-instantaneous nature of the interaction. A typical scan is placed in Figure \ref{fig:atmosphere}(c) to provide a weak-field reference for later analysis of the power-law response. When the delay between the spectrally split pulses is large enough, the spectrally separated pulses can be considered as non-interacting and perform conventional SHG within their own frequencies, while SHG by frequency mixing between the pulses is very weak or non existent [see Figure \ref{fig:atmosphere}(d)]. However, for short delays, the pulses temporally overlap leading to SHG by frequency mixing between the pulses yielding a broader SHG spectrum, as would be expected from a second order nonlinear process. 
\par
\begin{figure}[t]
\centering\includegraphics[width=\linewidth]{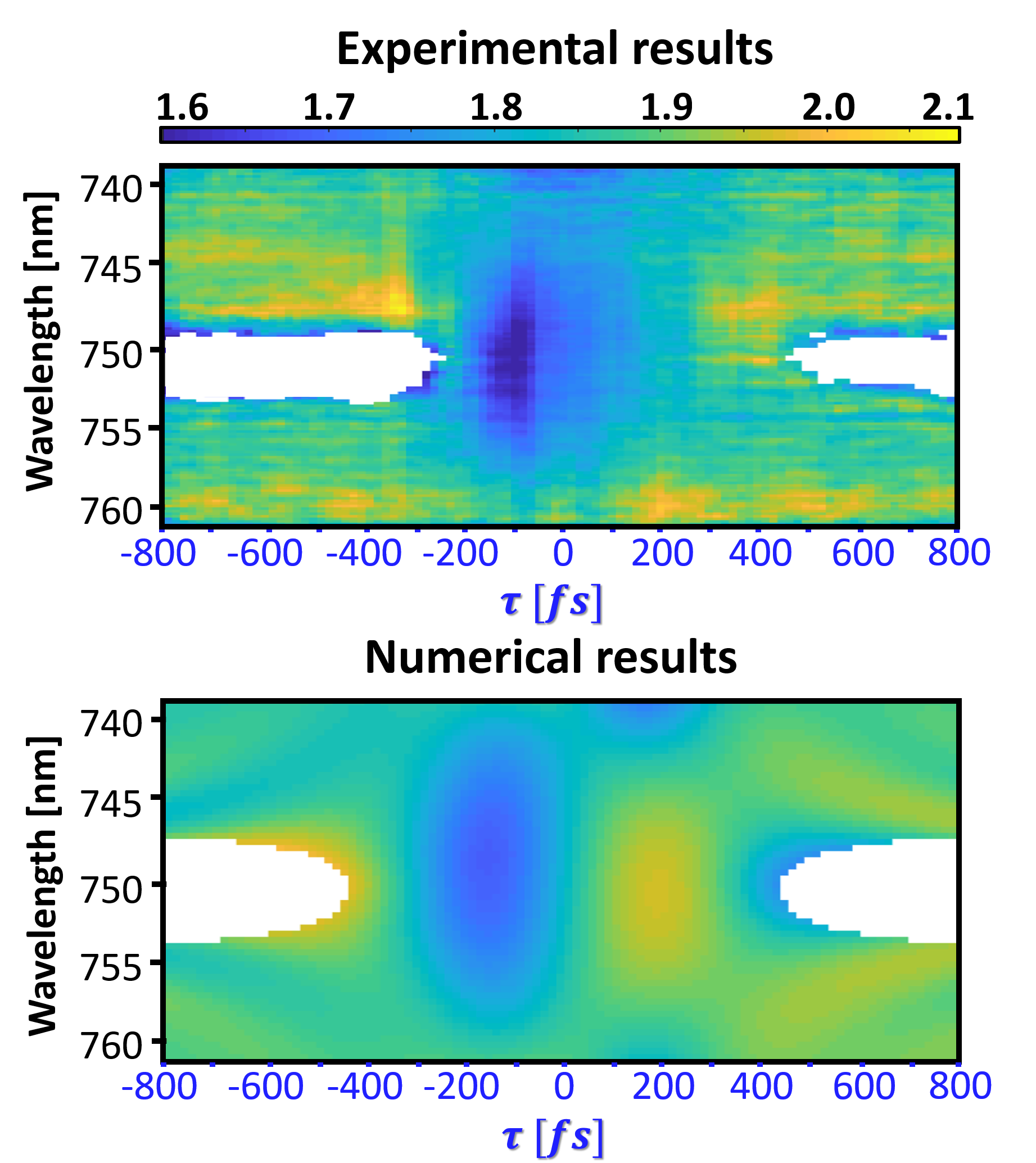}
 \caption{Nonlinear non-instantaneous response of SH in gold NSs. Map of the power-law response $P\left(\lambda,\tau\right)$ (color-coded) as a function of the delay between the pulses (x-axis) and wavelength (y-axis). (a) Experimental data : For a 100 fs negative delay, we observe that the nonlinear response decreases to 1.6 when shorter wavelengths arrive first (negative $\tau $), while in the case of swapping the order of arrival, such that longer wavelengths are first to arrive (positive $\tau $), this effect is absent. (b) Simulation results of the resonant three-level model. The main feature of a decrease in nonlinear response is predicted, as well as an overall well agreement with experiment. We note that both images have the same color bar. \label{H-shaped}}
\label{fig:H-shape}
	\end{figure}

 The experiment is repeated for different intensities to produce a pixel-by-pixel map showing the nonlinear power-law as a function of wavelength and slope value in terms of delay [see Figure \ref{fig:H-shape}(a)]. Appearing as white background in these plots, areas in which the SHG intensities are weak, such that they are comparable with the noise level  (1/100 for simulations), are considered irrelevant for power-law characterization, and therefore are cut out of the data. In our results, we see that unlike a conventional nonlinear crystal where the perturbative picture can be robustly modeled as $I^{P}$ (experimentally verified, BBO), we find the power-law response shows a complex, dynamical nature, which depends on delay and wavelength such that $ P\rightarrow P\left(\lambda,\tau\right) $. We point out three key features we observe in experiment: The nonlinear response is not symmetric in terms of delays, meaning that the order of arrival of the different pulses, each containing different spectral components, affects the power-law response for SHG. Furthermore, although peak intensity is higher for very short delays, the power-law response decreases to a minimal value of 1.6, which is significantly smaller than the conventional quadratic response expected in a second order process. Moreover, for zero delay, where the pulse is Fourier transform-limited (TL), peak intensity is at its highest yet the nonlinear response is not the strongest and is only $\sim $ 1.75. In a broader view, we observe that for small delays the nonlinear response is dynamical and sensitive to the order of arrival of pulses (sign of $\tau$). When shorter wavelengths are first to arrive, the nonlinear response decreases, gradually recovering as the delays shorten and fully recovers after 100 fs with longer wavelengths arriving first. Interestingly, for small delays, where the pulses overlap and peak intensity is at its highest, the power-law response is lower, suggesting the two photon picture is no longer suitable for describing the interaction process for these intensities. 
\par Our attempts to model the experimental results with the nonlinear oscillator model and its perturbative solution fails to predict the observed properties, such as the decrease in the SH power-law response. However, we find that modeling the nonlinear response as a resonant three-level,
can effectively predict the non-instantaneous properties emerging in experiment. These include, for the intermediate strength regime, a decrease in nonlinearity originating from the higher orders in the perturbative expansion. This attribute is not included in the nonlinear oscillator model, with susceptibility prescribed by Miller's rule \cite{boyd2003nonlinear}, where the nonlinearity is monotonically increasing when considering higher orders in the perturbative approach.
\par
\begin{figure}[h!]
\centering\includegraphics[width=\linewidth]{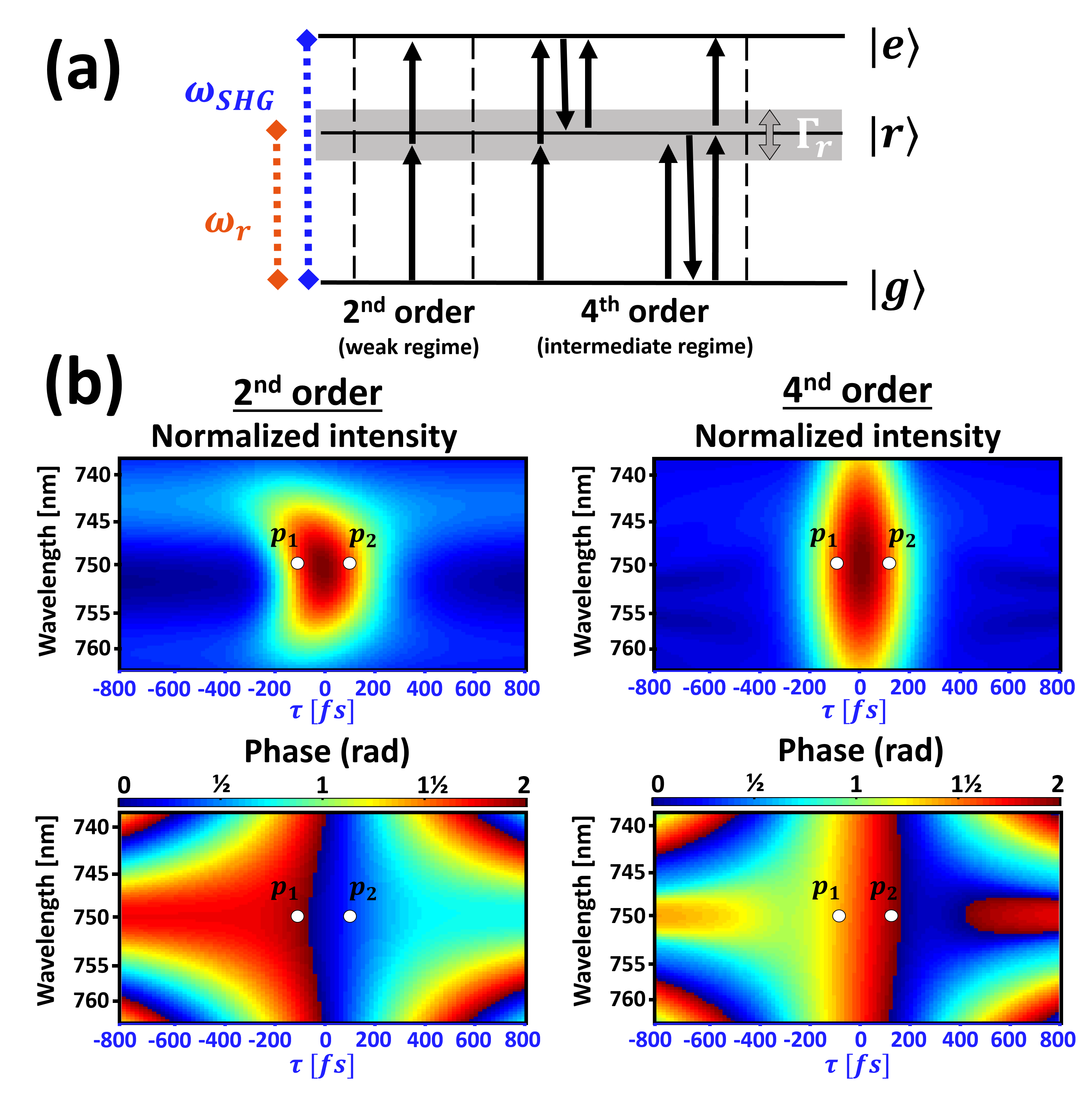}
 \caption{The resonant three-level system perturbative solution and analysis beyond \nth{2} order contributions. (a) Energy level diagram with second and fourth order interactions illustrated. $\omega_r$ and $\Gamma_r$ are the resonant frequency and linewidth of the nonlinear system, accordingly. Arrows represent the coupling of levels by the electric field, where opposite pointing arrows correspond to a conjugated electric field. (b) Second and fourth order intensity and phase plots, with intensities each normalized separately for clarity of the intensity structure. $p_1$ and $p_2$ are points later referred with further analysis.\label{level-system}}
\label{fig:level-system}
	\end{figure}

 In our model, we ascribe the intermediate level as the LSPR frequency. The excited state as the SH excitation frequency and ground state as the relaxed state of the system. We note that in the limit of a highly detuned resonant level, our solution approaches the known instantaneous solution found in non-resonant nonlinear optics \cite{boyd2003nonlinear}. By perturbative expansion, we  solve up to the fourth order. In the case of a sub-half octave spanning bandwidth, due to energy conservation, third order terms do not contribute to the excitation. Including the fourth order enables to capture intermediate field strength effects that impact the power-law response. We note that in our calculation of the fourth order pathways, terms which can be interpreted as cascaded processes with an effective coupling coefficient are included. The equations we obtain from this analysis are as follows
\begin{widetext}
\begin{subequations}
\begin{equation}
I_{SHG}\left(\omega\right) \propto {\left| \bar{\kappa}^2 E^{\left(2\right)} + \bar{\kappa}^{4} E^{\left(4\right)}\right|}^2 
\end{equation}
\begin{equation}
E^{\left(2\right)} \left(\omega \right)=\int{\frac{E\left(\omega-\Omega\right) \cdot E\left(\Omega\right)}{\omega_{r}-\Omega+i\Gamma_{r}}d\Omega}
\end{equation}

\begin{equation*}
E^{\left(4\right)}\left(\omega\right)= \int{\frac{E\left(\Omega_{2}\right)E\left(\Omega_{3}\right)E\left(\Omega_{4}\right)E\left(\omega-\Omega_{2}-\Omega_{3}-\Omega_{4}\right)}{\left(\omega_{r}-\Omega_{2}+i\Gamma_{r}\right)\left(\omega-\left(\Omega_{4}+\omega_{r}+i\Gamma_{r}\right)\right)\left(\Omega_{2}+\Omega_{3}-\omega-i\Gamma_{Au}\right)}d\Omega_2\Omega_3\Omega_4}
\end{equation*} 
\begin{equation}
+\int{\frac{E\left(\Omega_{2}\right)E\left(\Omega_{3}\right)E\left(\Omega_{4}\right)E\left(\omega-\Omega_{2}-\Omega_{3}-\Omega_{4}\right)}{\left(\omega_{r}-\Omega_{2}+i\Gamma_{r}\right)\left(\omega-\left(\Omega_{4}+\omega_{r}+i\Gamma_{r}\right)\right)\left(\Omega_{2}+\Omega_{3}-i\Gamma_{Au}\right)}d\Omega_2\Omega_3\Omega_4} 
\end{equation}
\end{subequations}
\end{widetext}
Where $E^{\left(2\right)}$ and $E^{\left(4\right)}$ are the second and fourth order contributions of the perturbative expansion. $\Omega_{2,3,4}$ are frequencies integrated over in the multiphoton process illustrated as arrows showing interaction pathways in Figure \ref{level-system}(a). $\omega_{r}$ is the measured LSPR frequency [see Figure 1(b)]. $\bar{\kappa}$, a parameter indicating coupling strength of the modes with the electric field that should depend on geometry and free electron contribution in such a process, was fitted to give the best agreement with the data. $ E\left(\Omega\right)=E_0 \cdot e^{i\phi\left(\Omega\right)}$ is the electric field, where $E_0$ and $ \phi\left(\Omega\right) $ are the measured electric field strength and spectral phase, accordingly. In our experiment, the electric field strengths are in the order of 0.1 $[\frac{V}{nm}]$. The linewidths are introduced into the calculation as suggested by Ref. \cite{weisskopf1930v,bebb1966multiphoton}, where $ \Gamma_{r}$ is the linewidth of the NS array acquired by transmission measurements and $\Gamma_{Au} $ is the collision rate related to the electron’s mean free path for a gold NS approximated according to Ref. \cite{cai2009optical}.
It is insightful to point out that $ E^{\left( 4\right)} $ is decomposed into two terms, each describing a different excitation process as illustrated in the pathway energy diagram in Figure \ref{fig:level-system}(a). For simplicity, their coupling strengths were approximated to be equal. Our simulation results are presented in Figure \ref{fig:H-shape}(b), predicting the main feature of a decrease in nonlinearity, in an asymmetric temporal response and with the same time scale. In our simulations, we see that time scales are determined by the LSPR’s linewidth and Au collision rate. The results are relatively robust in terms of positive/negative detuning, where the sign of the detuning mostly affects the wavelength of the power-law's minimum. The simulations also predict that TL pulses do not yield the strongest power-law response and show other similar features observed in experiment. We see noticeable differences between the experiment and predicted results such as a 5\% decrease in both $\tau =200$ and $\tau =-130$ relative to the simulations. We attribute these differences to the spectral response of the NS’s compared to the ideal Lorentzian shape of the three-level system, which become more significant for higher order excitation leading to a decreased power-law response.
\par When including terms beyond \nth{2} order contributions in the three-level model, the next order's contribution is, as expected, initially small and becomes dominant with increased power. Surprisingly, the power-law response decreases, for the intermediate regime, due to phase difference with higher order excitations, effectively suppressing the excitation which corresponds to SHG. The intensity and phase of the \nth{2} and \nth{4} orders are plotted in the Figure \ref{fig:level-system}(b). Although the direct ascertainment of the nonlinear behaviour depends on both intensity and phase, observing the areas which are differently phased gives insight to the effect emerging in the nonlinear response. To unravel this connection, we perform analysis in the complex plane. For a specific wavelength and spectral shape, we plot the intensity (magnitude) and phase (angle) of the \nth{2} and \nth{4} order contributions separately and their sum. We further repeat this calculation with increased intensities.
\par
\begin{figure}[h!]
\centering\includegraphics[width=\linewidth]{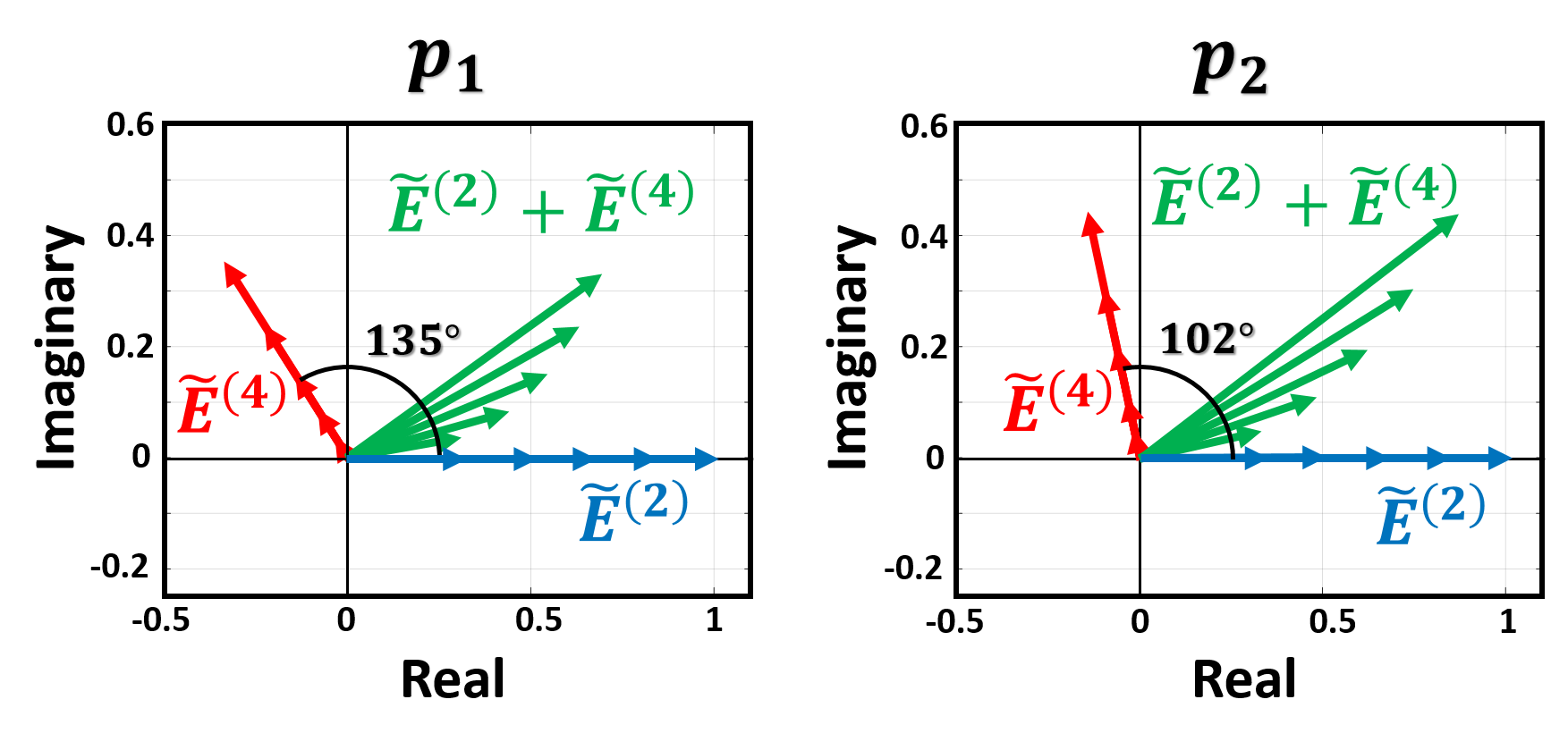}
 \caption{The importance of relative phase in strong field SHG. Complex plane representation of the second, fourth and summed contributions for different intensities showing how relative phase and intensity affects the nonlinear response. These are plotted for a specific wavelength and delays marked $p_1$ and $p_2$ in \ref{fig:level-system}(b). The local evolution of the magnitude of the summed contribution would be the power-law response exhibited for each process. Each plot is rotated such that $E^{(2)}$ is pointing in the real direction, and normalized by $E^{(2)}$'s maximal value. By changing the order of arrival of the excitation pulses (negative to positive $ \tau $), the relative phase decreases and the intensity drives the system past its minimal nonlinear response to yield a stronger response.}
\label{fig:arrows}
    \end{figure}

 We show two examples from our analysis [Figure \ref{fig:arrows}]. For $p_1$, the phase between the \nth{2} and \nth{4} order leads to destructive interference resulting in a nonlinear response which is smaller than quadratic. For $p_2$, the smaller phase difference is sufficient for the \nth{4} order intensity to drive the combined sum past its minimal point towards \nth{4} order nonlinearity resulting in an effectively restored quadratic nonlinearity.
In order to fully appreciate the impact of the relative phase on the power-law's response, we verified and discuss here the ideal cases of minimal and maximal phase differences. In the case of no phase difference, the combined contributions' magnitude would be, for low intensities, comparable to the \nth{2} order contribution, increasing similarly. For high intensities, it would increase similarly to the \nth{4} order contribution. Here, the rate of increase in magnitude, which is proportional to the power-law response, would increase \textit{monotonically}. The transition from \nth{2} to \nth{4} order response still occurs when a phase difference exists. However, for the intermediate regime, the increase in magnitude is exchanged for rotation, resulting in a weaker power-law response. In the opposite case of a $180\degree$ phase difference, the contributions completely interfere destructively, leading to smaller than quadratic response for the intermediate regime. Here, the power-law response would initially decrease, passing its minimal point, monotonically approaching towards \nth{4} order nonlinearity. In our simulations, these phase differences in the SH excitation are the origin for the dynamical response and we believe describes the dynamical nonlinear response of the differently phased excitation of the LSPR found by experiment.
\par To conclude, by coherent control we experimentally unravel the ultrafast dynamical, non-instantaneous nonlinear response in resonant NSs. We see a pronounced decrease in the power-law response when shorter wavelengths arrive prior to longer wavelengths. We show experimentally that TL pulses do not yield the strongest nonlinearity when approaching strong field regime. We introduce a novel theoretical approach, which capture effects beyond the weak-field two-photon regime and describes our experimental observations in an intuitive picture. Since this model is based on the general behaviour of any system with a resonant level it may be used to describe other nonlinear dynamical observations such as the non-instantaneous SHG measured near a resonant intraband transition \cite{guo2001ultrafast} or the observed deviations of LSPR dephasing times from the ideal harmonic oscillator model \cite{anderson2010few}. Furthermore, it could be applied to study other systems such as resonant excitonic and polaritonic couplings, SHG in van der Waals materials \cite{lin2019quantum}, or for coherent two-photon luminescence in NSs \cite{remesh2018phase}. 

\setlength{\parskip}{13pt} We acknowledge funding from the European Research Council (ERC) under the European Union’s Horizon 2020 research and innovation program (Grant Agreement MIRAGE 20-15) and by the Israel Science Foundation (ISF) through Grant No. 1433/15.

\bibliography{bibliography_NLsrr.bib}
\end{document}